\documentclass{article}
\usepackage{spconf,amsmath,graphicx}
\usepackage[hidelinks]{hyperref}
\usepackage{bm}
\usepackage{comment}
\usepackage{algorithm}
\usepackage{algorithmic}
\usepackage{booktabs}
\usepackage{cite,color}
\usepackage[normalem]{ulem}
\usepackage[bottom]{footmisc}
\usepackage[position=top]{caption}
\usepackage[acronym]{glossaries}
\makeatletter
\renewcommand\fs@ruled{\def\@fs@cfont{\bfseries}\let\@fs@capt\floatc@ruled
  \def\@fs@pre{\hrule height.8pt depth0pt \kern2pt}%
  \def\@fs@post{\kern2pt\hrule\relax\kern-1mm}%
  \def\@fs@mid{\kern2pt\hrule\kern2pt}%
  \let\@fs@iftopcapt\iftrue}
\makeatother
\makeglossaries 
\newcommand{\INPUT}{\item[\textbf{Input:}]}
\newcommand{\OUTPUT}{\item[\textbf{Output:}]}

\title{How Reliable Are Predicted MOS for Reproducing\\Human System-Level Preferences in Speech Enhancement?}

\name{\shortstack[c]{Nahomi Kusunoki$^1$, Tsubasa Ochiai$^2$, Naohiro Tawara$^2$, Marc Delcroix$^2$, \\
 Naoyuki Kamo$^2$, Tetsuji Ogawa$^1$, Shoko Araki$^2$}}
\address{
  $^1$Waseda University, Japan \ \ \ $^2$NTT, Inc., Japan
}
\begin{document}
\ninept
\maketitle
\begin{abstract}

We investigate whether predicted Mean Opinion Scores (MOS) can reliably support system-level comparisons of speech enhancement (SE) methods by introducing \emph{system-level preference accuracy (SPA)}.
Although MOS prediction models are widely used to evaluate SE systems, their performance is typically assessed by correlation with human-rated MOS, 
which does not guarantee agreement on which system is better.
SPA addresses this gap by directly evaluating 
whether predicted and human-rated MOS yield the same system preferences. 
Using SPA, we systematically evaluate three settings: single prediction models, ensembling, and domain adaptation.
Through experiments, 
SPA varies substantially across single prediction models, from 9.4\% to 76.8\%.
Even the best 
model disagrees with human judgments in approximately 23\% of system comparisons.
Ensembling yields only limited improvement, while domain adaptation tends to substantially improve SPA in the closed condition but brings only modest gains in the more practical open condition, where neither the target systems nor the speakers are known.
 These results suggest that SPA can reveal errors correlation-based evaluation alone does not expose, and that predicted MOS alone can lead to unreliable conclusions in practical SE system comparison.

\end{abstract}
\section{Introduction}
\label{sec:intro}

Subjective evaluation has been increasingly adopted in recent speech enhancement (SE) research to evaluate the perceptual audio quality of the processed speech. Among subjective evaluation methods, human-rated Mean Opinion Score (MOS)~\cite{ITUTP800,ITUTP808,ITUTP835} is the most commonly used metric~\cite{sach2025p,zhang2025lessons}. 
For a pair of systems, human-rated MOS scores are compared via a statistical significance test (e.g., a paired t-test) to determine which system performs better, or whether the difference is not significant. 

However, conducting subjective evaluation requires human annotation, which is time-consuming and financially costly. Therefore, SE research has been increasingly using MOS prediction models~\cite{wang2025improving,stahl2025distillation,reddy2021dnsmos,mittag2021nisqa,saeki2022utmos,ragano2024scoreq} as a more practical alternative\cite{sun2025efficient,asai2026geneses}. 
These models predict the MOS that listeners would assign, and the relative superiority between systems is then typically judged based on the system-level averaged predicted scores. 
Despite this use, these models are conventionally evaluated using correlation-based metrics (e.g., linear correlation coefficient (LCC) and Spearman's rank correlation coefficient (SRCC)~\cite{dodge2008concise}), which do not directly measure their ability to determine whether one SE system is superior to another\cite{saeki2022utmos,reddy2021dnsmos,mittag2021nisqa,cooper2022generalization}. 
For example, a high SRCC value does not imply that every system pair is correctly ordered, and it can remain high even when some pairwise orderings are reversed. 
Therefore, it is not necessarily clear whether MOS prediction models can reflect the pairwise system preferences of human-rated MOS.

To evaluate the reliability of SE system comparison, it is desirable to directly examine to what extent pairwise system preferences based on predicted MOS reproduce those obtained from human-rated MOS.
Thus, in this paper, we introduce system-level preference accuracy (SPA)\footnote{The code for computing SPA will be made publicly available.}, which quantifies such agreement as the rate of matching pairwise system preferences: one system is superior, the other is superior, or the two are comparable.
We compute SPA with bootstrap sampling to quantify the uncertainty from data variability.

Using SPA, we comprehensively analyze the reliability of pairwise system preferences based on predicted MOS in two stages. 
We first examine their impact on SPA when a single MOS prediction model is used.
Next, we examine the impact on SPA of two approaches reported to improve MOS prediction performance: model ensembling~\cite{saeki2022utmos,kunevsova2023ensemble} and domain adaptation~\cite{cooper2022generalization,do2023resource}.

The key findings of this study can be summarized as follows.
\begin{enumerate}
\setlength{\itemsep}{0pt}   
\setlength{\parsep}{0pt}    
\item To the best of our knowledge, we conducted the first comprehensive analysis of whether pairwise system preferences based on predicted MOS agree with those based on human-rated MOS, by introducing a novel SPA metric, which directly measures the extent of this agreement.
\item We evaluated six MOS prediction models on two different SE tasks, including widely used DNSMOS\cite{reddy2021dnsmos}, NISQA\cite{mittag2021nisqa}, and UTMOS\cite{saeki2022utmos}. SPA mainly ranged from 30.8\% to 76.8\%. Even for the best-performing model and dataset, the pairwise system preference was still incorrect in about 23\% of cases, raising concerns for practical use in SE system comparisons.
\item We evaluated ensembles combining multiple predictors via averaging or majority voting. SPA did not consistently improve over the best single model. Even in the oracle system combination, the pairwise system preference was still incorrect in about 17\% of cases, suggesting that using multiple predictors does not necessarily resolve the reliability concerns.
\item  
We evaluated domain adaptation under closed and open conditions, where target systems and speakers are seen or unseen.
The experimental results suggest that domain adaptation can enable reliable judgments in the closed condition for some models and datasets, achieving an SPA of up to 98.8\%, but reliability issues remain in the more practical open condition.

\end{enumerate}

\vspace{-5mm}
\textbf{\section{Related Work}
\label{sec:related}}

Some studies on MOS prediction models have introduced a training loss that aligns utterance-level preferences between predicted and human-rated MOS \cite{wang2023mospc,wang2025urgent,wang2026urgentmos}, i.e., which of the two utterances receives the higher score.
These studies have also reported utterance-level preference accuracy as an evaluation metric, which measures whether predicted and human-rated MOS agree in judging the preference between a pair of utterances.
However, utterance-level and system-level preferences tend to exhibit different properties, and in SE research, systems are typically evaluated and compared based on their system-level 
(i.e., averaged) 
predicted MOS scores. 
Therefore, system-level preference accuracy is better aligned with this practice and thus better suited to our goal of evaluating the reliability of MOS prediction models in SE research.

\vspace{4mm}
\section{System-level Preference Accuracy}
\vspace{-1mm}

We formulate SPA as a three-way classification problem that directly
evaluates the agreement between pairwise system preferences based
on predicted and human-rated MOS, i.e., one system is superior (win), the
other system is superior (lose), or the two systems are comparable (tie).
Additionally, we adopt the bootstrap sampling to account for the
variability of the evaluation dataset 
and to quantify the statistical uncertainty in the SPA estimate.

Algorithm~\ref{al:spa_multi} describes the computational procedure of the SPA. 
Let $\mathbf{x}_i^s$ and $m_i^s$ denote the output of SE system $s$ for utterance $i$ and its corresponding human-rated MOS, respectively. By inputting $\mathbf{x}_i^s$ to the MOS prediction model $\mathrm{pMOS}(\cdot)$, the predicted MOS score $\hat{m}_i^s$ is obtained.
$\mathcal{U} = \{ 1, \ldots, N^{\text{U}} \}$ and $\mathcal{S} = \{ 1, \ldots, N^{\text{S}} \}$ denote the indices of utterances and SE systems, where $N^{\text{U}}$ and $N^{\text{S}}$ are their total numbers.
$\mathcal{C}$ denotes the set of all possible system pairs of $\mathcal{S}$.
$\mathrm{q}_{0.025}(\cdot)$ and $\mathrm{q}_{0.975}(\cdot)$ denote 2.5\% and 97.5\% quantiles.

SPA, defined as $\bar{P}_b$ in line~11, is the proportion of system pairs in $\mathcal{C}$ whose preference agrees between the predicted and human-rated MOS, ranging from 0 (no pairs agree) to 1 (all pairs agree).
In lines 5 and 7, the difference of average MOS for a system pair is computed for the human-rated and predicted  scores, respectively.
In lines 6 and 8, these differences are converted into a win/lose/tie preference via the tie-aware sign function $\mathrm{sign}_\epsilon(\cdot)$, defined as $\mathrm{sign}_\epsilon(z) = \mathrm{sign}(z)\,\mathbf{1}(|z|>\epsilon)$ with tie threshold $\epsilon \ge 0$, where $\mathrm{sign}(\cdot)$ and $\mathbf{1}(\cdot)$ denote the sign and indicator functions.
SPA is computed by averaging the agreements of these preferences in line~11.
With bootstrap sampling (lines 1--2), SPA ($\bar{P}_b$) is computed over utterances resampled with replacement in each iteration; the mean of the resulting distribution is reported as the overall SPA score, $P(\mathcal{C})$ (line 13), along with its 95\% confidence interval (CI), $\mathrm{CI}_{0.95}(\mathcal{C})$ (line 14).

\begin{algorithm}[!t]
\footnotesize
\caption{SPA Computation with Bootstrap Sampling}
\label{al:spa_multi}
\begin{algorithmic}[1]
\INPUT
Dataset:
$\mathcal{D} = \{ \mathcal{D}_{s} \}_{s \in \mathcal{S}}$,
where
$\mathcal{D}_{s} = \{ (\mathbf{x}_{i}^{s}, m_{i}^{s}) \}_{i=1}^{N^{\mathrm{U}}}$;
MOS prediction model:
$\mathrm{pMOS}(\cdot)$;
Number of bootstrap iterations:
$N^{\mathrm{B}}$.
\OUTPUT
Mean of the SPA
$P(\mathcal{C})$
and 95\% CI
$\mathrm{CI}_{0.95}(\mathcal{C})$.
\FOR{$b = 1$ to $N^{\mathrm{B}}$}
  \STATE $\mathcal{U}_b \gets$
  sample $N^{\mathrm{U}}$ indices from $\mathcal{U}$ with replacement
  \FOR{each system pair $c = (A,B)$ in $\mathcal{C}$}
  \STATE Take $\{(\mathbf{x}_i^s, m_i^s)\}_{i\in\mathcal{U}_b}$ from $\mathcal{D}_s$, $s\in\{A,B\}$
  \STATE $\Delta^{\mathrm{human}}_{A,B}
  \gets
  \frac{1}{N^{\mathrm{U}}}
  \sum_{i\in\mathcal{U}_b}
  (m_i^A-m_i^B)$
  \STATE $v_{\mathrm{human}}
  \gets
  \mathrm{sign}_{\epsilon}
  (\Delta^{\mathrm{human}}_{A,B})$
  \STATE $\Delta^{\mathrm{pred}}_{A,B}
  \gets
  \frac{1}{N^{\mathrm{U}}}
  \sum_{i\in\mathcal{U}_b}
  \left(\mathrm{pMOS}(\mathbf{x}_i^A)
  -
  \mathrm{pMOS}(\mathbf{x}_i^B)\right)$
  \STATE $v_{\mathrm{pred}}
  \gets
  \mathrm{sign}_{\epsilon}
  (\Delta^{\mathrm{pred}}_{A,B})$
  \STATE $P_c \gets \mathbf{1}(v_{\mathrm{pred}}=v_{\mathrm{human}})$
  \ENDFOR
  \STATE $\bar{P}_b
  \gets
  \frac{1}{|\mathcal{C}|}
  \sum_{c\in\mathcal{C}}
  P_c$
\ENDFOR
\STATE $P(\mathcal{C})
\gets
\frac{1}{N^{\mathrm{B}}}
\sum_{b=1}^{N^{\mathrm{B}}}
\bar{P}_b$
\STATE $\mathrm{CI}_{0.95}(\mathcal{C})
\gets
[
\mathrm{q}_{0.025}(\{\bar{P}_b\}_{b=1}^{N^{\mathrm{B}}}),
\mathrm{q}_{0.975}(\{\bar{P}_b\}_{b=1}^{N^{\mathrm{B}}})
]$
\end{algorithmic}
\end{algorithm}

\section{Approaches for Improving MOS Prediction}
\label{ssec:improvement_methods}

\vspace{-1mm}
\subsection{Ensemble of Multiple MOS Prediction Models}
\vspace{-1mm}

We investigate whether ensembling multiple MOS prediction models can improve the reliability of SE system comparison.
SE research sometimes uses predicted MOS scores from multiple models. 
In this case, 
most SE papers judge superiority by considering these scores comprehensively.
We formalize this practice as an ensemble of MOS prediction models and evaluate its effect on SPA.

In this paper, we consider two ways of constructing it: majority voting (MV) and averaging (Ave).
For ensembling, we assume that $L$ MOS prediction models 
$\{\mathrm{pMOS}_l(\cdot)\}_{l=1}^{L}$ are available.
In MV, the preference $v_{\mathrm{pred}}^{(l)}$ is computed by $\mathrm{pMOS}_l(\cdot)$ for each $l$, following the same procedure as lines 7--8 of Algorithm~\ref{al:spa_multi}.
The ensembled preference score $v_{\mathrm{pred}}$ is then determined by majority voting selecting the most frequent value in $\{ v_{\mathrm{pred}}^{(l)} \}_{l=1}^{L}$.
In Ave,
the ensembled predicted MOS score $\tilde{m}_i^s$ is computed by averaging $L$ predicted MOS scores as 
$\tilde{m}_i^s = (1/L) \sum_{l=1}^{L} \mathrm{pMOS}_l(x_i^s)$. 
SPA is then computed by replacing $\mathrm{pMOS}(\cdot)$ in line 7 of Algorithm~\ref{al:spa_multi} with the averaged score.
Note that MV does not compute the predicted MOS scores, and thus the correlation-based metrics cannot be evaluated.

Prior studies have explored ensembling multiple sub-models or learners to improve prediction performance~\cite{saeki2022utmos,kunevsova2023ensemble}. We instead ensemble independently developed MOS prediction models (e.g., DNSMOS, NISQA, and UTMOS) because they are widely used for evaluation in SE research, which allows us to examine the reliability of SE system comparisons using multiple prediction models.

\vspace{-2mm}
\subsection{Domain Adaptation of MOS Prediction Model}
\label{sssec:domain_adaptation}
\vspace{-1mm}

We investigate whether domain adaptation can help predicted MOS better reproduce human system-level preferences.
Domain adaptation finetunes the MOS prediction model on a limited amount of target-domain data, mitigating the mismatch between the training and evaluation domains and thereby improving its prediction performance.
We evaluate its effect on SPA under two adaptation conditions: closed and open.
The closed condition assumes that speech and its corresponding human-rated MOS from the target speakers and systems are available for adaptation.
The open condition is more practical because they are typically unknown in advance, but generalization to such unseen conditions is generally difficult.

\vspace{-1mm}
\section{Speech Quality Evaluation Experiments}
\vspace{-1mm}
\subsection{Experimental Setup}
\label{ssec:ex_set_single_spa}

\textbf{Datasets: } We used two recently released datasets:  the evaluation set of the CHiME-7 UDASE dataset~\cite{leglaive2023chime} (5 systems, 7 speakers, 640 samples) and the blind test set of the URGENT24 dataset~\cite{zhang2025lessons} (23 systems, 177 speakers, 6900 samples).
While CHiME-7 UDASE contains real-world conversational speech recorded in multi-speaker reverberant environments, URGENT24 covers a variety of speech distortion types.
Following ITU-T P.835\cite{ITUTP835} and P.808\cite{ITUTP808}, CHiME-7 UDASE and URGENT24 conducted subjective listening tests, from which we use OVRL MOS and MOS scores, respectively.

\noindent
\textbf{Models: }We used six representative MOS prediction models: DNSMOS\cite{reddy2021dnsmos}, NISQA\cite{mittag2021nisqa}, UTMOS\cite{saeki2022utmos}, Distill-MOS\cite{stahl2025distillation}, SCOREQ\cite{ragano2024scoreq}, and UniVERSA-Ext\cite{wang2025improving}, using their publicly released implementations \cite{Distill-MOS,Urgent2026,NISQA-code,UTMOS-code,DNSMOS-code,SCOREQ-code}.
DNSMOS is trained primarily on DNS Challenge speech, while NISQA and SCOREQ are trained on VoIP speech.
UTMOS is trained mainly on TTS speech, while Distill-MOS and UniVERSA-Ext are trained across diverse domains.

\noindent
\textbf{Evaluation metrics: }We used system-level SRCC (sys-SRCC) and SPA as evaluation metrics. 
SPA is computed with bootstrap sampling ($N^{\mathrm{B}}=1000$) unless otherwise specified. 
To determine the tie threshold $\epsilon$, we computed the minimum
significant difference (MSD) using Tukey's honestly significant
difference (HSD) test, obtaining $0.122$ for CHiME-7 UDASE and
$0.131$ for URGENT24.
Moreover, since OVRL MOS and MOS are averaged over eight listeners,
their resolution is $1/8 = 0.125$.
Motivated by these observations, we adopted $\epsilon = 0.125$ as
the tie threshold.
Although the assumptions of Tukey's HSD test may not strictly hold here, we leave a more principled choice of $\epsilon$ to future work.

\noindent
\textbf{Adaptation  details: }We evaluated the effect of domain adaptation based on the cross-validation (CV) scheme.
Since it is preferable for SPA to compare two systems using the same MOS prediction model, we defined each CV subset in terms of a speaker (or speaker group) and a system pair, rather than a single system.
For CHiME-7 UDASE, we used every combination of speaker and system pair as a subset. For URGENT24, since the number of speakers is much larger, we divided the speakers into two groups and instead used every combination of speaker group and system pair as a subset.

Using this CV scheme, we obtained an adapted MOS prediction model for each speaker (or speaker group) and system-pair subset, and used it to compute the predicted MOS of each utterance in the corresponding subset. For each system pair, the predicted MOS scores were aggregated across the corresponding speaker subsets and fed into Algorithm~\ref{al:spa_multi} to compute SPA.

For the adaptation data, we defined two evaluation conditions: closed and open.
In the closed condition, all subsets other than the evaluation subset were candidates for the adaptation data.
In the open condition, subsets sharing the same speaker or system as the evaluation subset were excluded from the candidates for the adaptation data.
For each fold, the adaptation data was constructed
via stratified sampling from the candidate subsets to satisfy the number of adaptation samples $n$.

As an exception, since the enhanced signals differ substantially from the mixture, we handle the mixture subsets in a system-closed manner to avoid unintended out-of-domain conditions in the open condition: all mixture subsets other than the one used for evaluation are included in the candidate subsets.

For learning-rate tuning, we randomly selected one CV fold in the closed condition and partitioned its adaptation set into 90\% for adaptation and 10\% for development. We tuned the learning rate based on the LCC scores of the development set.

\vspace{-1mm}
\subsection{Results of Individual MOS Prediction Models}
\label{subsec:result_single}

\begin{table}[!t]
\centering
\caption{System-level SRCC (``sys-SRCC'') and SPA [\%] with 95\% CIs. \#M indicates number of models used for ensemble. For Ave (Best), \#M = 2 on CHiME-7 UDASE and \#M = 3 on URGENT24. }
\vspace{-2mm}
\label{tab:main_results}
\setlength{\tabcolsep}{2pt}
\setlength{\aboverulesep}{0pt}
\setlength{\belowrulesep}{0pt}
\renewcommand{\arraystretch}{1.55}
{\small
\begin{tabular}{l|c|cc|cc}
\toprule
& & \multicolumn{2}{c|}{\textbf{CHiME-7 UDASE}} & \multicolumn{2}{c}{\textbf{URGENT24}} \\
\cline{3-4} \cline{5-6}
\textbf{Model} & \textbf{\#M} &
\textbf{sys-SRCC} & \textbf{SPA} &
\textbf{sys-SRCC} & \textbf{SPA} \\
\midrule
UniVERSA.  & 1 & 0.900 & $62.0^{+8.0}_{-12.0}$ & 0.938 & $76.8^{+4.2}_{-3.7}$ \\
SCOREQ    & 1 & 0.900 & $65.7^{+4.3}_{-15.7}$ & 0.926 & $73.6^{+3.1}_{-3.6}$ \\
Distill-MOS  & 1 & 0.700 & $49.4^{+0.6}_{-9.4}$  & 0.755 & $73.1^{+4.0}_{-3.5}$ \\
UTMOS     & 1 & 0.200 & $9.4^{+0.6}_{-9.4}$   & 0.902 & $72.8^{+3.9}_{-4.0}$ \\
NISQA     & 1 & 0.200 & $40.9^{+9.1}_{-0.9}$  & 0.702 & $67.1^{+3.3}_{-3.4}$ \\
DNSMOS    & 1 & -0.100 & $30.8^{+9.2}_{-0.8}$ & 0.556 & $52.9^{+3.7}_{-3.4}$ \\
\midrule
MV (All)   & 6 & - & $42.5^{+7.5}_{-2.5}$ & - & $70.4^{+2.5}_{-3.2}$ \\
Ave (All)  & 6 & 0.300 & $40.8^{+9.2}_{-0.8}$ & 0.958 & $79.8^{+4.0}_{-3.9}$ \\
MV (Best)  & 3 & - & $49.8^{+0.2}_{+0.2}$ & - & $78.6^{+1.5}_{-2.8}$ \\
Ave (Best) & 2 / 3 & 0.900 & $80.6^{+9.4}_{-20.6}$ & 0.969 & $83.1^{+3.5}_{-4.0}$ \\
\bottomrule
\end{tabular}
\vspace{-0mm}
}
\end{table}

We first evaluate the reliability of individual MOS prediction models in Table~\ref{tab:main_results}.
SPA values, which are computed with bootstrap sampling, are shown in the form $x_{a}^{b}$, where $x$ is the overall SPA score, and $a$ and $b$ indicate the differences of $x$ from the lower and upper bounds of the 95\% CI, respectively, as computed in Algorithm~\ref{al:spa_multi}.
Incidentally, the CI of MV (Best) on CHiME-7 UDASE is $[50\%, 50\%]$; accordingly, $a$ and $b$ in Table~\ref{tab:main_results} are set to $0.2$.
 We observe that even when the MOS prediction models differ substantially in terms of sys-SRCC, their SPA scores do not always differ to the same extent. 
 This suggests that SPA captures an aspect of a model's reliability for SE system comparison that is not necessarily reflected in sys-SRCC, supporting its use as a complementary metric.

On CHiME-7 UDASE, SPA mainly ranged from $30.8\%$ to $65.7\%$, indicating that the evaluated MOS prediction models made incorrect pairwise system preferences for roughly one-third to two-thirds of the system pairs. 
When
accounting for the 95\% CIs, the lower bounds fall within
roughly $30\%$ to $50\%$ across models. 

On URGENT24, we confirm that SPA was generally higher than on CHiME-7 UDASE, but the highest SPA among the evaluated models was still $76.8\%$. 
Even the best-performing model can lead to an incorrect conclusion in roughly $23\%$ of comparisons on average.
These results imply that a single prediction model can often be insufficient for reliably comparing SE systems in practice.

\subsection{Results of Multiple MOS Prediction Models (Ensemble)}
We next investigate whether combining multiple MOS prediction models improves the reliability of SE system comparison. 
Table~\ref{tab:main_results} (lower part) reports the  sys-SRCC and the overall
SPA scores with their 95\% CIs for two ensemble strategies: Ave and MV. ``All'' denotes the ensemble of all six models, while ``Best'' denotes the combination that achieved the highest SPA among all possible subsets of the six models (i.e., combinations of size 2 to 6), showing the upper-bound (oracle) performance achievable through ensembling.

On CHiME-7 UDASE, Ave (All) achieved an SPA of only $40.8\%$, below that of the best single model ($65.7\%$), suggesting that combining all six models is not necessarily helpful, probably because the individual models themselves were unreliable on this dataset. However, Ave (Best) achieved a substantially higher SPA of $80.6\%$, indicating that a considerable improvement over the best single model is possible if the best-performing combination could be identified. In contrast, MV (Best) notably underperformed Ave (Best), reaching only $49.8\%$, probably because majority voting is more vulnerable when the individual models' SPA is low.

On URGENT24, where the individual models performed relatively better than on CHiME-7 UDASE, Ave (All) achieved an SPA of $79.8\%$, exceeding the best single model's SPA of $76.8\%$. Ave (Best) further improved this to $83.1\%$. In contrast, MV (Best) reached only $78.6\%$, again underperforming Ave (Best). Across both datasets, Ave consistently outperformed MV, suggesting that averaging predicted scores may offer a more reliable basis for SE system comparison than comparing multiple individual judgments.

Overall, even in the oracle best case, SPA reached only $83.1\%$ on URGENT24 and $80.6\%$ on CHiME-7 UDASE, a level that can still be insufficient for reliably comparing SE systems. These results suggest that using multiple MOS prediction models, whether through averaging or majority voting, does not necessarily resolve the reliability concerns raised for individual models.

\vspace{-1mm}
\subsection{Results of Domain Adaptation for MOS Prediction Models}

\begin{table}[!t]
    \centering
    \caption{SPA [\%] of adapted models for closed and open conditions.}
    \vspace{-2mm}
    \label{tab:ft-preference}
    \setlength{\tabcolsep}{3pt}
    \setlength{\aboverulesep}{0pt}   
    \setlength{\belowrulesep}{0pt}   
    \renewcommand{\arraystretch}{1.3} 
    \scalebox{0.8}{
    \begin{tabular}{@{}c|cc|cc|cc|cc|cc@{}}
        \toprule
        \multicolumn{11}{c}{\textbf{CHiME-7 UDASE}} \\
        \midrule
        & \multicolumn{2}{c|}{\textbf{UniVERSA.}}
        & \multicolumn{2}{c|}{\textbf{SCOREQ}}
        & \multicolumn{2}{c|}{\textbf{UTMOS}}
        & \multicolumn{2}{c|}{\textbf{Distill-MOS}}
        & \multicolumn{2}{c}{\textbf{NISQA}} \\
        \cmidrule(lr){2-3}\cmidrule(lr){4-5}\cmidrule(lr){6-7}\cmidrule(lr){8-9}\cmidrule(lr){10-11}
        n & closed & open & closed & open & closed & open & closed & open & closed & open \\
        \midrule
        0    & \multicolumn{2}{c|}{$62.0$} & \multicolumn{2}{c|}{$65.7$} & \multicolumn{2}{c|}{$9.4$} & \multicolumn{2}{c|}{$49.4$} & \multicolumn{2}{c}{$40.9$} \\
        100  & $83.0$ & $71.7$ & $92.4$ & $86.1$ & $81.2$ & $50.6$ & $88.7$ & $40.4$ & $62.1$ & $40.4$ \\
        300  & $96.7$ & $81.4$ & $98.2$ & $85.9$ & $87.7$ & $64.0$ & $98.8$ & $44.2$ & $85.7$ & $40.3$ \\
        \midrule\midrule
        \multicolumn{11}{c}{\textbf{URGENT24}} \\
        \midrule
        & \multicolumn{2}{c|}{\textbf{UniVERSA.}}
        & \multicolumn{2}{c|}{\textbf{SCOREQ}}
        & \multicolumn{2}{c|}{\textbf{UTMOS}}
        & \multicolumn{2}{c|}{\textbf{Distill-MOS}}
        & \multicolumn{2}{c}{\textbf{NISQA}} \\
        \cmidrule(lr){2-3}\cmidrule(lr){4-5}\cmidrule(lr){6-7}\cmidrule(lr){8-9}\cmidrule(lr){10-11}
        n & closed & open & closed & open & closed & open & closed & open & closed & open \\
        \midrule
        0    & \multicolumn{2}{c|}{$76.8$} & \multicolumn{2}{c|}{$73.6$} & \multicolumn{2}{c|}{$72.8$} & \multicolumn{2}{c|}{$73.1$} & \multicolumn{2}{c}{$67.1$} \\
        200  & $76.2$ & $76.3$ & $78.4$ & $77.2$ & $79.0$ & $77.6$ & $67.9$ & $69.5$ & $61.4$ & $60.3$ \\
        1000 & $78.4$ & $77.4$ & $81.3$ & $83.3$ & $82.0$ & $77.8$ & $75.2$ & $69.6$ & $61.8$ & $64.8$ \\
        \bottomrule
    \end{tabular}
    }
    \vspace{2mm}
\end{table}

Finally, we examine the effect of domain adaptation on SPA.
Table~\ref{tab:ft-preference} shows the overall SPA scores over bootstrap resamples after adaptation with a small amount of data.
Due to space constraints, the CIs of SPA are omitted here.
We use 100 and 300 samples for CHiME-7 UDASE, and 200 and 1000 samples for URGENT24.
Here, $n=0$ denotes the SPA of the original, non-adapted model (as in Table~\ref{tab:main_results}).
DNSMOS is excluded from this experiment because its ONNX-based distribution made adaptation nontrivial to implement.

On CHiME-7 UDASE, under the closed condition, SPA improved substantially for every model with more adaptation samples.
SCOREQ and Distill-MOS reached particularly high SPA scores of $98.2\%$ and $98.8\%$ after adaptation.
These results indicate that, with a suitable model, domain adaptation can make system comparisons highly reliable when the target systems and speakers are known in advance, though this is often not the case in practice.
Under the more practical open condition, SCOREQ maintained a relatively high SPA of $85.9\%$, still an incorrect judgment in about 14\% of cases, whereas Distill-MOS instead dropped to 44.2\%.
This indicates that adaptation does not consistently generalize to unseen systems and speakers, and also its benefit can vary greatly depending on the model.

In contrast, on URGENT24, the effect of adaptation was much more limited overall under both the closed and open conditions.
We can confirm that some models showed modest improvements, while others instead performed worse than without adaptation.
We conducted additional experiments with different learning rates, but this degradation was not resolved under any conditions.
This is probably due to the diversity of speech distortion types and the large number of systems in URGENT24.

Overall, these results indicate that domain adaptation can make system comparisons highly reliable under the closed condition, but only for some models and datasets, and its benefit is much more limited under the more practical open condition.
This suggests that domain adaptation alone is unlikely to be sufficient to guarantee reliable system comparisons.

\subsection{Analysis of SPA Error Types}

\begin{figure}[!t]
    \centering\includegraphics[width=0.45\textwidth]{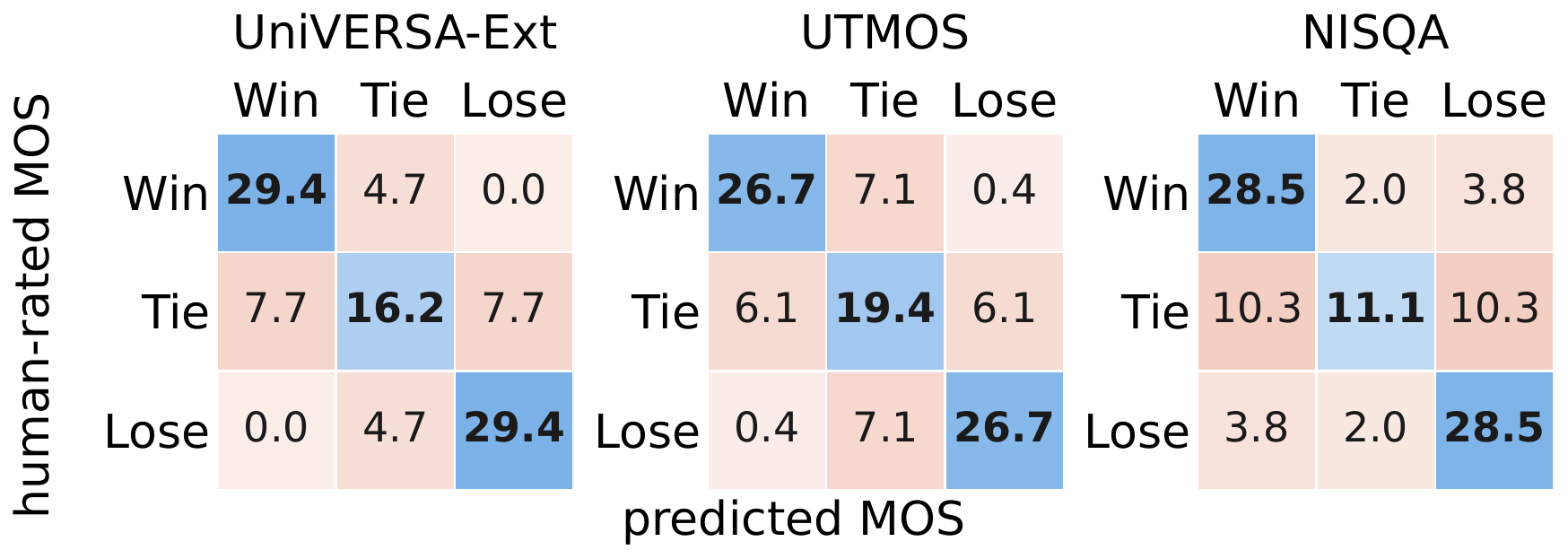}
    \vspace{1mm}
    \caption{Confusion matrix of human-rated MOS vs.\ predicted MOS on URGENT24.}
    \label{fig:urgent24_confusion}
    \vspace{-5mm}
\end{figure}

To further analyze the types of errors in SPA, we examine the confusion pattern between human-rated and predicted pairwise system preferences (i.e., win, tie, lose) on the URGENT24 dataset, as shown in Fig.~\ref{fig:urgent24_confusion}, without bootstrap sampling.
Due to space constraints, the figure shows only three models.
For every ordered pair of systems\footnote{Considering both $(A,B)$ and $(B,A)$ avoids an arbitrary dependence on which system is labeled first.}, we determine its human-rated and predicted win/tie/lose decisions, and each cell of the matrix gives the percentage of such pairs with the corresponding human-rated decision vs. predicted decision.
The diagonal reflects SPA agreement, while the off-diagonal reflects two error types: win/lose reversals and tie-related confusions.

Win/lose reversals, i.e., cases where a human-rated win was predicted as a lose or vice versa, were consistently small across all models, ranging from 0.0\% to 7.6\%.
In contrast, tie-related confusions, i.e., cases where either the human-rated or predicted judgment was a tie, were consistently large, ranging from 20.8\% to 40.4\%.
These accounted for the large majority of SPA errors.
This suggests that predicted MOS is relatively robust for detecting sufficient performance gaps between systems, but unreliable for fine-grained comparisons between systems with closely matched scores.
For most of the six evaluated models, 
false differentiation errors (predicting a win or lose for a human-rated tie) occurred more frequently than false equivalence errors (predicting a tie for a human-rated win or lose): for NISQA, the former totaled 20.6\% (10.3\% × 2), while the latter totaled 4.0\% (2.0\% × 2).
This indicates that predicted MOS more often mistakes genuinely comparable systems for different ones (win or lose), which should be considered a critical type of error.

\section{Conclusion}

In this paper, we introduced system-level preference accuracy (SPA), a metric that directly evaluates whether predicted MOS can reproduce system-level preferences based on human-rated MOS.
Our results showed that a single MOS prediction model is often insufficient for reliably determining pairwise system preferences, even when it achieves high correlation with human-rated MOS.
Ensembling and domain adaptation both offered partial improvements, but neither guaranteed reliable judgment: even the best-performing combination of models left about 17\% of judgments incorrect, and while domain adaptation could make SE system comparisons highly reliable under the closed condition for some models and datasets, its benefit was far more limited under the more practical open condition.
These findings show that SPA can reveal disagreements between predicted and human-rated pairwise system preferences that correlation-based evaluation alone does not expose, supporting its use as a complementary metric for evaluating the reliability of SE system comparison.
Future work will investigate more principled approaches to setting the tie threshold in SPA, including statistical-significance-based decision rules.

\section{Acknowledgements}

We thank the organizers of the URGENT 2024 challenge for providing supplementary data (e.g., speaker information) beyond the publicly released dataset. 

\bibliographystyle{IEEEbib}
\bibliography{strings,refs}

\end{document}